# Non-invasive lipid quantification of living microalgal cultures with digital holographic microscopy

CATHERINE YOURASSOWSKY,[1,3] RENAUD THEUNISSEN,[2] JÉRÔME DOHET-ERALY,[1] AND FRANK DUBOIS[1,4]

[1]*Microgravity Research Centre, Université libre de Bruxelles, 50 Avenue Franklin Roosevelt, 1050 Bruxelles, Belgium*
[2]*Bio, Electro and Mechanical Systems: Embedded Electronics, Université libre de Bruxelles, 50 Avenue Franklin Roosevelt, 1050 Bruxelles, Belgium*
[3]*Catherine.Yourassowsky@ulb.be*
[4]*Frank.Dubois@ulb.be*

**Abstract:** Some microalgae store large amounts of neutral lipids inside lipid droplets. Since these lipids can be used to produce biodiesel in a sustainable way, research is developing on fast non-destructive methods to quantify and monitor the amount of lipids in microalgal cultures. In this paper, we have developed with digital holographic microscopy a fast quantitative method to assess the evolution of the lipid content inside the diatom *Phaeodactylum tricornutum* living cells. The method uses a specific processing of recorded hologram sequences based on the refocusing capability of digital holographic microscopy. Each lipid droplet volume is evaluated inside the cells on representative samples of the culture. We have validated the method thanks to correlative quantitative phase contrast–fluorescence imaging and extrapolated it to larger calibrated spherical refractive particles, to demonstrate the flexibility of the method.

## 1. Introduction

Research is increasing for the production of microalgal lipids for, in particular, a sustainable biodiesel production [1–6]. Indeed, microalgae store high neutral lipid content in lipid bodies, also called lipid droplets (LDs), which can be transformed into biodiesel. Some microalgae like the diatom *Phaeodactylum tricornutum* are easy to culture, fast growing and have a fatty acid composition appropriate for biodiesel production [4,5,7].

In this field, it is therefore important to have fast non-destructive methods to quantify and monitor the amount of lipids in microalgal cultures. It is particularly significant for the analysis of the best culture conditions, for the best strain selection, and for the determination of the optimal day of harvest.

The quantification of the microalgal lipids is routinely performed by solvent extraction, gravimetric estimation, and lipid profiling with chromatography. However, these processes are destructive and can cause a loss of lipid quantity. They need large samples and are time-consuming [6].

Fluorescent techniques like fluorescence microscopy [7–10] or laser scanning confocal microscopy [11,12] work with reduced sample amount, and allow single-cell analysis to study LDs in detail, but they are invasive by the need to stain the lipids of the cells with specific lipophilic fluorescent dyes. Moreover, those dyes present some limitations like photobleaching and variable penetration inside the cells due to the cell wall barrier [6].

Raman microscopy [6,13] was used to perform non-destructive, label-free quantification of the lipid accumulation in microalgal cultures thanks to the detection of lipid-specific Raman spectral peaks. However, this technique is slow and invasive, as it needs a high-powered light source. The coherent anti-Stokes Raman scattering (CARS) microscopy was successfully used to visualize and quantify the microalgal LDs [6,13–15]. Moreover, it allows to separate the lipid-specific signals from the excited chlorophyll fluorescence and to analyze LD formation [15]. It is a label-free microscopy technique but requires multiple laser sources that may affect the cell physiology, making this technique invasive [10].

In order to improve the monitoring of the evolution of microalgal LDs, quantitative phase contrast imaging can be performed with digital holographic microscopy (DHM), a non-invasive method that was used to quantify the LDs in murine macrophages cultures [16].

Campos *et al.* [17] used DHM to perform the monitoring of adipocyte differentiation during the culture. The quantification of the lipids was monitored on the cultures thanks to the evolution of the optical path difference.

Kim *et al.* [18] developed the optical diffraction tomography (ODT), able to establish the 3D refractive index distribution of an individual cell from multiple 2D holographic images recorded with different angles of illumination. This non-invasive method was used to perform the quantification of lipid content in hepatocytes [19] and in individual microalgal cells [20]. However, it is necessary to record from 200 up to 300 holograms with a sequential angle scanning thanks to a digital micro-mirror device to get the information in a small field of view. It is therefore difficult to perform the high-throughput analysis of a microalgal culture.

Guo *et al.* [21] developed the time-stretch quantitative phase microscopy to classify microalgal cells cultivated in two different conditions in a high-throughput way. The optical set-up is relatively complex and individual LDs are not identified.

In this paper, we develop a fast quantitative method to assess the lipid content inside the diatom *Phaeodactylum tricornutum* living cells, thanks to a specific processing of hologram sequences recorded with a DHM. This method is partly based on a previous method that we developed in DHM to characterize spheroid transparent objects thanks to their capacity to concentrate the light, providing an intensity peak like a usual converging lens [22]. It is linked to the specific DHM post-recording capability of numerical refocusing that allows the in-depth scanning of the light intensity distribution. This method was already applied to cells in order to determine different cell characteristics allowing, for example, the classification of living and dead cells in cultures [22].

In a similar way, L. Miccio *et al.* [23] and P. Memmolo *et al.* [24] used the formation of a focal-spot (the intensity peak) from red blood cells to perform blood diagnosis and to detect abnormal cells thanks to focal-spot analysis. They also demonstrated the imaging capability of red blood cells that behave like adaptive liquid-lenses with tunable focal length. These properties could be used to develop new bio-imaging tools.

As the intracellular microalgal LDs are usually spheroid and with a higher refractive index than the external medium, we have developed the present method for the detection and the size evaluation of the LDs inside the living cells of microalgal cultures. It allows the assessment, in a non-invasive way, of the amount of lipid inside many individual living cells. This provides statistically representative data on the lipid content of the culture. We have applied it to cultures of the diatom *Phaeodactylum tricornutum*.

## 2. Material and methods

### 2.1 Optical set-up

A reduced spatial coherence multimode digital holographic microscope (DHM) is implemented to estimate the volume of LDs inside microalgal cells. As described in our patent [25] and publications [26,27], the instrument combines the DHM with the fluorescence mode. The complete holographic information and the fluorescence signals emitted by the sample are recorded sequentially with the same camera. This multimode DHM allows to perform correlative quantitative phase contrast-fluorescence microscopy. The overlay of the quantitative phase contrast images and the fluorescent images is a powerful tool to identify specific cells or cellular elements. In this study, it is used to identify and localize inside the algal cells the fluorescent LDs, stained with BODIPY 505/515, in the corresponding phase image and to validate the method.

The multimode DHM is schematized in Fig. 1. In this instrument, the DHM is based on a Mach–Zehnder interferometer working in transmission with a partially spatial coherent illumination that is described in [28]. Thanks to this illumination the raw holograms can be directly visualized on the computer screen. Indeed, with fully coherent illumination, the direct image is usually too noisy to be interpreted by a direct viewing [29,30].

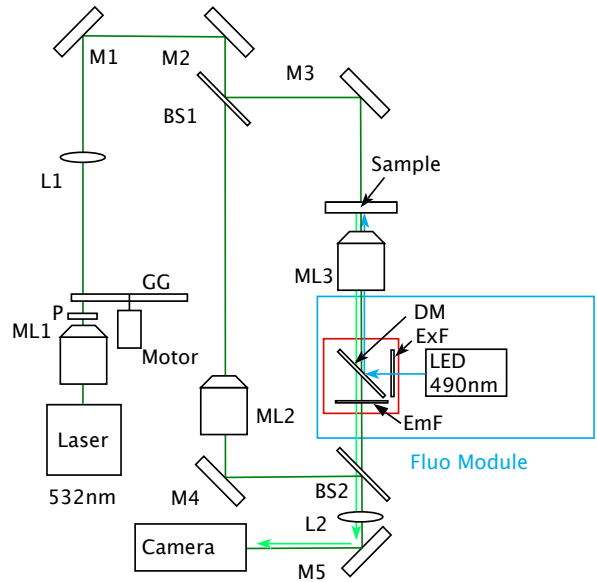

Fig. 1. Optical set-up of the multimode DHM. BS1–BS2, beam splitters; DM, dichroic mirror; EmF, emission filter; ExF, excitation filter; GG, rotating ground glass; L1–L2, lenses; M1–M5, mirrors; ML1–ML3, microscope lenses; and P, polarizer.

For the DHM mode, the laser source is a Cobolt Samba $^{TM}$ CW 532 nm DPSSL. The microscope lenses ML2 and ML3 are oil immersion Leica 100×, numerical aperture (NA) 1.3, or Leica 40×, NA 0.60. The camera is a CCD camera Hamamatsu ORCA ER, with a CCD array of 1344×1024 pixels cropped to 1024×1024 pixels with a pixel size of 6.45 µm × 6.45 µm. The fields of view are 66×66 µm$^2$ (100× magnification) for the LD experiments, and 171×171 µm$^2$ (40× magnification) for the experiments on silica particles. The fluorescent light source is a CoolLED pE-100 system with a LED at 490 nm coupled with a fluorescence filter set appropriate for BODIPY 505/515.

*2.2 Holograms and fluorescence signals, recording with the multimode DHM*

The *Phaeodactylum tricornutum* unlabelled culture samples were injected in the micro-channel of an ibidi µ-Slide I 0.2 Luer (ibidi$_®$) and placed on the DHM stage. Sequences of one hundred digital holograms were recorded during the translation of the DHM stage, perpendicularly to the optical axis.

For the fluorescence imaging of stained LDs, the laser source was switched off. The LED source at 490 nm was switched on for the BODIPY 505/515 fluorescence excitation and the fluorescence signals were recorded by the CCD camera. The corresponding holograms were recorded with the same camera, using the partially spatial coherent source at 532 nm while the LED source was switched off.

*2.3 Culture conditions*

The *Phaeodactylum tricornutum* strain (UTEX 646) was obtained from the Culture Collection of Algae at the University of Texas at Austin and was grown at 15 °C in 250 mL flasks containing 100 mL sterilized filtered seawater with 90 µl of f/2 medium [31] with the salinity of 34 ppt. The cultures were illuminated with a fluorescent lamp at 100 µmol photons.m$^{-2}$.s$^{-1}$ on a 12:12 hours light/dark cycle. To avoid bacterial contamination and for nutrient purposes, we also added 30 µL of antibiotic solution made by mixing 1 mL of purified water (milli-Q) with 35 mg of Penicillin G sodium salt and 35 mg of Streptomycin sulphate salt, 15 µL of vitamins (2.96×10$^{-7}$ M of thiamine, 2.05×10$^{-9}$ M biotin, and 3.69×10$^{-10}$ M cyanocobalamin), and 30 µL of silica (Na$_2$SiO$_3$) solution. Those flasks were daily shaken.

*2.4 Sample preparation for fluorescence imaging*

The LDs of *Phaeodactylum tricornutum* living cells were stained with the lipophilic fluorescence dye BODIPY 505/515 (Invitrogen) following the protocol described in L. He *et al.* [7]. The BODIPY 505/515 was diluted in anhydrous dimethyl sulfoxide (DMSO) to obtain a final concentration of 100 mM. 10 µL of this solution was mixed with 1 mL of the *Phaeodactylum tricornutum* culture, kept in a dark place and let settle during 5 minutes at room temperature (25 °C) to ensure the fluorescent labelling. The samples were injected in the micro-channel of an ibidi µ-Slide I 0.2 Luer and placed on the DHM stage.

*2.5 Calibrated particles for the method validation and extrapolation*

To validate the holographic method and to extrapolate it up to larger LDs as found inside adipocytes, we also tested it on calibrated monodisperse spherical silica particles of 7.82 μm diameter with a standard deviation (SD) of 0.31 μm (micro particles GmbH). As those particles are significantly larger than the LDs of *Phaeodactylum tricornutum*, we implemented in the DHM microscope objectives Leica 40×, NA 0.60.

*2.6 Processing to compute the LD sizes inside the microalgal cells*

In order to compute the size of the LDs inside the cells, different processing steps were applied to the recorded holograms of the culture samples.

2.6.1. Extraction of the digital holographic information

This step is achieved using the usual Fourier transform method in off-axis holography that consists in computing the discrete 2D Fourier transformation of the hologram $h_m(s,t)$, where $s,t$ are integers ($s,t = 0,...,N-1$) and $m$ the hologram label, isolating one of the side lobes, shifting its center into the origin of the Fourier plane, and computing the inverse Fourier transformation. It results the complex amplitude $g_{0,m}(s,t)$, where the index 0 indicates that the complex amplitude is computed in the recorded plane. As previously mentioned, the holograms were cropped to 1024×1024 pixels; we kept this size for all the images in the further processing.

2.6.2. Propagation up to the focal planes

The holograms are not necessarily recorded in the best focus plane of the microalgal cells in the sample. Therefore, it is necessary to propagate $g_{0,m}(s,t)$ up to the best focus planes of the microalgal cells by using the usual propagation equation. The propagation over a distance $d$ is given by:

$$g_{d,m}(s',t') = \exp\{jkd\} F^{-1} Q[-\lambda^2 d] F^{+1} g_{0,m}(s,t), \qquad (1)$$

where $s',t' = 0,...,N-1$ are integers, $j = \sqrt{-1}$, $\lambda$ is the wavelength, $k = 2\pi/\lambda$ is the wavenumber, $F^{\pm 1}$ represent the direct (+) and inverse (−) discrete Fourier transformations and $Q[a]$ is a quadratic phase factor depending on the real parameter $a$:

$$Q[a] = \exp\left\{\frac{jka}{2N^2\Delta^2}(U^2 + V^2)\right\}, \qquad (2)$$

with $\Delta$ the sampling distance in the input plane, and $U,V$ the discrete spatial frequencies varying from 0 to $N-1$.

According to the refocusing criteria [32], and as the microalgal cells are mainly phase objects, the best focus distances are found by searching the propagation distance $d$ for which the quantity

$$a_d = \sum_{s',t'=0}^{N-1} |g_{d,m}(s',t')| \qquad (3)$$

is maximum.

In general, it is necessary to individually apply the refocusing criteria for each microalgal cell isolated within a region of interest in the complex amplitude images $g_{d,m}$ [33]. However, in these particular experiments, the sedimentation process located most of the microalgal cells on the bottom wall of the experimental slide in such a way that the refocusing criteria can be applied, in a very good approximation, to the full complex amplitude image $g_{d,m}$. According to the requested accuracy for the next processing step, the maximum of $a_d$ was obtained by performing refocusing steps of 0.01 μm.

2.6.3. Recording of a large number of holograms and elimination of the phase and modulus amplitude artifacts

We recorded, for each sample, 100 images of the experimental slide with different lateral translation positions, in order to record representative sets of microalgal culture. Those sequences of holograms were also applied to remove the phase background and the permanent amplitude modulus defects by performing averaging processes over all the recorded sequences [33]. After those processes, it resulted sequences of corrected $g_{0,m}(s,t)$ complex amplitudes.

As we used for the LDs 100× magnification microscope objectives, with a high numerical aperture (NA = 1.3), the depth of focus (typically 0.25 μm) is very small, and the lateral motions of the slide to record the sequence of images may disturb the focus. Therefore, it is necessary to perform the automated digital refocusing in each $g_{0,m}(s,t)$ as described here above.

2.6.4. Detection of the LDs by digital holographic refocusing and determination of the diameter of each LD

The procedure for detecting and measuring individual LDs is illustrated in Fig. 2. It is based on the fact that every LD may be seen as a lens, hence focusing light in a forward plane.

We assume that the refractive indices $n_1$ and $n_2$, respectively within a droplet and outside the cell (i.e., the culture medium) are close to each other, in such a way that the paraxial approximation can be used. Indeed, in this case, $n_1$ is approximately 1.47 (vegetable oils [20]) and $n_2$ is about 1.34 (sea water). The optical effect of the cytoplasm, which consists in a thin layer located between the LDs and the cell membrane, is weak and may be therefore ignored. Under these hypotheses, we can consider that the LD introduces a quadratic phase factor, as a usual thin lens, to an incident plane wave, which is given by [34]

$$\exp\left\{-j\frac{k}{2r}(x^2+y^2)\right\}, \qquad (4)$$

where $r$ is the light focusing distance, i.e., the focal length of the lens, and $(x,y)$ are the position coordinates in the plane of the thin lens, with $(0,0)$ in the center of the lens. Moreover, the optical path difference emerging out of the droplet, between the center and the border of the droplet, is $D(n_1-n_2)$, where $D$ is the droplet diameter, which corresponds to a phase difference of $kD(n_1-n_2)$. This result is used to adjust the parameter $r$ in Eq. (4):

$$\exp\left\{-j\frac{k}{2r}\left[(x^2+y^2)\big|_{x=0,y=0}-(x^2+y^2)\big|_{x=D/2,y=0}\right]\right\}=\exp\left\{jkD(n_1-n_2)\right\}; \qquad (5)$$

thus $(D/2)^2/2r = D(n_1-n_2)$, which gives

$$r = D/[8(n_1-n_2)]. \qquad (6)$$

Consequently, the light focusing distance $r$ with respect to the LD focus plane is proportional to its diameter $D$ divided by the refractive index difference.

As the exact refractive index difference is not necessarily simple to determine in practice, it can be assumed that, for a given experimental situation, it is constant. It has also to be pointed out that the influence of the cytoplasm is weak. Therefore, one can model

$$br = D, \qquad (7)$$

where $b$ is a constant experimentally determined as explained below.

As described in 2.6.2, the complex amplitude has been first computed in the focus plane of the LDs. Then, the light focusing zone corresponding to each droplet is searched and detected by performing incremental digital reconstructions with Eq. (1). For that purpose, 120 reconstructions by steps of 0.05 µm are performed to detect the maximal local intensities. Rigorously, for every pixel, the maximal value of the intensity, among all the 120 reconstructed images, is kept. This reconstruction range and the requested sensitivity were determined after a preliminary check. This process provides an image with the highest intensity while reconstructing by 6 µm along the optical axis. For each pixel, the reconstruction distance corresponding to the kept value of the intensity is stored. As the droplets are focusing the light, the intensities of each focused spot are locally significantly larger than the average intensity of the reconstructed image. Therefore, applying a threshold process to the reconstructed image intensities allows the individual identification of the droplets. In practice, the average image intensity and the threshold are, respectively, set to the grey levels 32 and 150 for every image (on a 256 grey level scale). The threshold operation creates small zones around the maximum focusing intensities. However, the smallest ones are not relevant as they are due to perturbations that may happen. To eliminate them, a connectivity surface analysis is performed to reject the zones having an area lesser than 20 pixels. It results a collection of the zones of interest covering the individual focusing spots generated by the LDs [Fig. 2(c)]. It has to be emphasized that the detection of the focused spots isolated by a threshold provides an efficient way to segment the droplets. The next step consists in searching in each zone of interest the maximal value of the intensity and to read the corresponding reconstructing distance, which gives an estimation of the focalization distance $r$ of the LD. Thanks to the relationship $br = D$, the diameter of each LD is subsequently assessed.

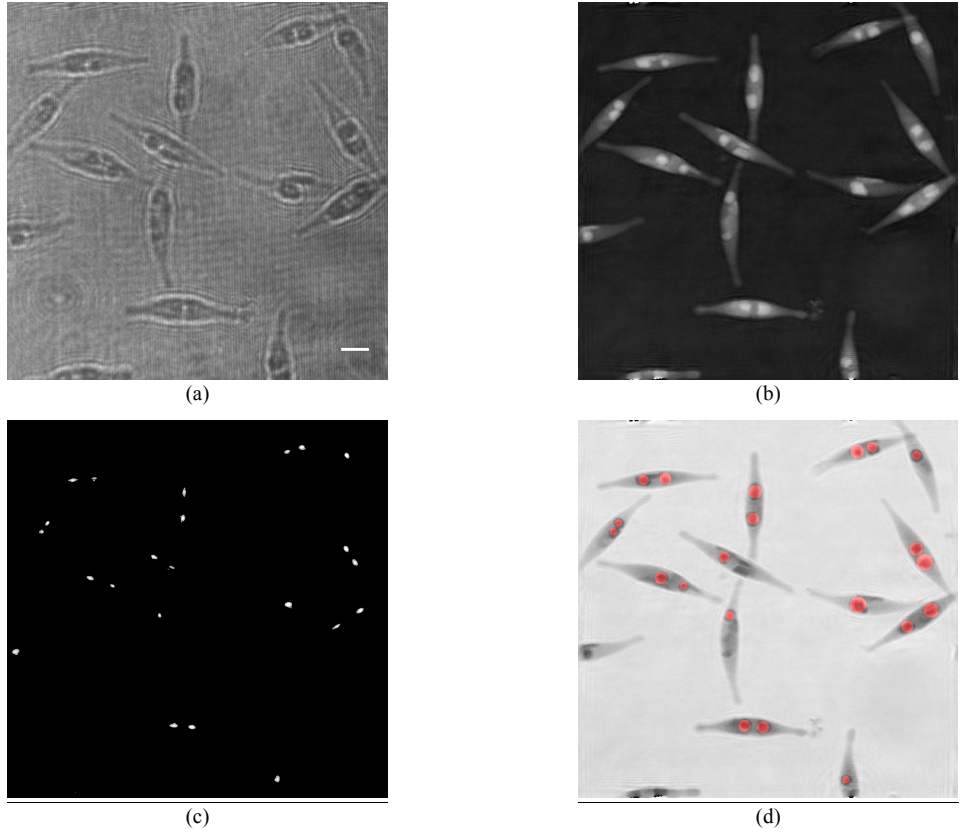

(a)                          (b)

(c)                          (d)

Fig. 2. Detection and size assessment of LDs inside cells of a *Phaeodactylum tricornutum* culture at 24 days. (a) Recorded hologram; (b) corresponding refocused phase image; (c) zones of highest intensities while reconstructing over a range of 6 μm along the optical axis; each zone corresponds to one detected LD; and (d) superimposition of the inverted phase [in (b)] with the computed disks representing the assessed LDs, in red; the radii of the red disks are computed according to the presented method. Scale bar = 5 μm.

In comparison with usual methods using images in the focal plane of LDs, this approach provides the huge advantage that LDs in contact give well-separated focusing spots. This is inherent to the method and largely simplifies the segmentation process of the LDs.

After this step, we dispose of a set of well-separated segmented spots individually identifying the LDs and the corresponding diameters. An image with disks centered in each focused zone and having the computed diameters, superimposed on the respective phase image, is created to illustrate the efficiency of the developed method [Fig. 2(d)]. The aforementioned phase background correction is particularly important for this purpose, since a non-flat phase background would induce a lateral bias in the determination of the position of the LDs.

The determination of the $b$ parameter is made as follows. This parameterization is specific for the sample type and is therefore kept for all the analyses of the same kind of samples. We consider images like in Fig. 2(d), with the red disks superimposed on the inverted phase images, and where red disks are displayed while varying the $b$ parameter. A visual assessment, with a random series of lipid droplets, then allows to accurately define a range of admissible $b$ values: by applying a sequence of $b$ values, one can clearly see, by superposition with the phase image, when the red disks are too small or too large, even for relatively small incremental values of $b$. We observed that the relative error $\overline{\Delta b/b}$ is about 0.1. It should be emphasized that the main goal of the proposed method is to provide a tool to study the evolution of LD sizes, which is not crucially impacted by the exact value of $b$, according that it must be kept identical for all the sequences of experiments.

An important aspect to determine is the smallest size of the detectable droplets. This limit finds its origin in the lateral resolution limit $0.61\lambda/\text{NA}$. Indeed, the focusing effect of the droplets imaged by digital holography is also impacted by the lateral resolution limit, by the way of a convolutional process with the point-spread function $p(x,y)$. Indeed, for an in-focus input amplitude $g_0(x,y)$, the output amplitude in the detection plane can be expressed by $g_0 \otimes p$, where $\otimes$ represents the convolution operation. This amplitude is propagated by a distance $r$, which corresponds to the focusing spot generated by the droplet; it results $R[d](g_0 \otimes p) = p \otimes R[d]g_0$, where $R[d]$ denotes the free space propagation operator. This indicates that the focused spot is convolved by the same point spread function as the input amplitude. Therefore, for droplets with size comparable to the resolution limit, we cannot expect to obtain the necessary increase in intensity by the focusing effect. In order to determine what could be the minimum size of detected droplets with our method, we simulated the modification to a plane wave transmitted through a droplet of diameter $D$ with inside and outside refractive indices of 1.47 (lipid) and 1.34 (water), respectively. We propagated it up to the best focalization plane, We convolved the resulting amplitude with point spread functions corresponding to different NA's and we measured the ratio $q$ between the maximum and the background intensities. We assumed that the droplet is detectable if $q$ is larger than 3, which corresponds to a realistic threshold. We studied two cases:

(1) $D$ set to 3 µm and decreasing NA down to get $q = 3$; we obtained NA = 0.6, which gives a ratio between $D$ and the resolution of 5.5; and
(2) NA set to 1.3 and decreasing $D$ to get $q = 3$; we obtained $D = 1.2$ µm, which gives a ratio between $D$ and the resolution of 4.8.

It can be concluded that the minimum size of a detected droplet is about 5 times the resolution limit. With our set-up for the LDs, that corresponds to $D = 1.25$ µm.

## 3. Results

### 3.1. Measurements

The results obtained on 4 series of different cultures at 24 days are given in Table 1. The number of detected LDs is relatively large that makes the method ideal for monitoring living cell cultures by a statistically significant way. We observe that the mean values of the LD volume are in the range 4.85–6.28 µm$^3$ at 24 culture days that is in good agreement with the values obtained by Wong *et al.* [12] with a laser scanning confocal microscope. The standard deviations are relatively important, which is also observed by Wong *et al.* [12]. It means that LD size distribution is widespread.

|  | Number of detected LDs | LD mean volume | Standard deviation |
|---|---|---|---|
| Serie 1 | 1302 | 4.85 µm$^3$ | 3.17 µm$^3$ |
| Serie 2 | 1422 | 6.28 µm$^3$ | 4.02 µm$^3$ |
| Serie 3 | 2753 | 5.57 µm$^3$ | 3.48 µm$^3$ |
| Serie 4 | 2667 | 5.56 µm$^3$ | 3.65 µm$^3$ |

Table 1. Assessment of Individual LD Size Inside Cells of 4 Series of Phaeodactylum tricornutum Cultures at 24 Days

### 3.2. Validation of the method by correlative quantitative phase contrast–fluorescence imaging

The fluorescence channel of the multimode DHM allowed to validate the identification of the LDs in the phase images and the formation of the intensity peak by each LD. Figure 3 shows an example of results on a culture of 24 days.

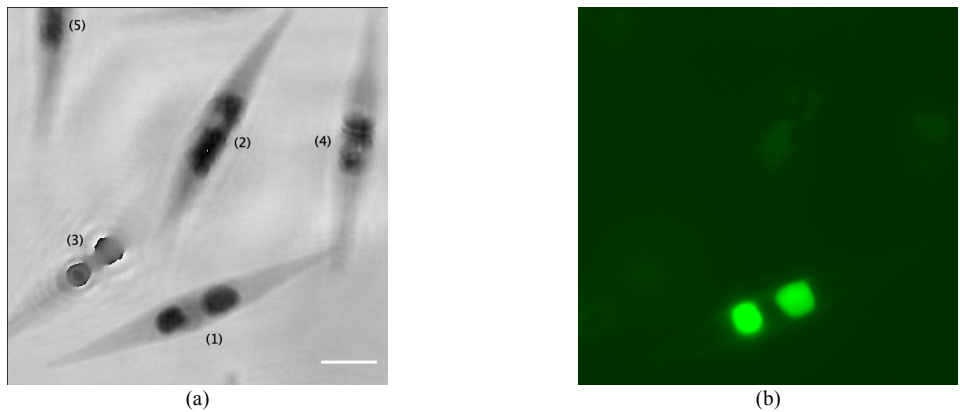

(a)                                                    (b)

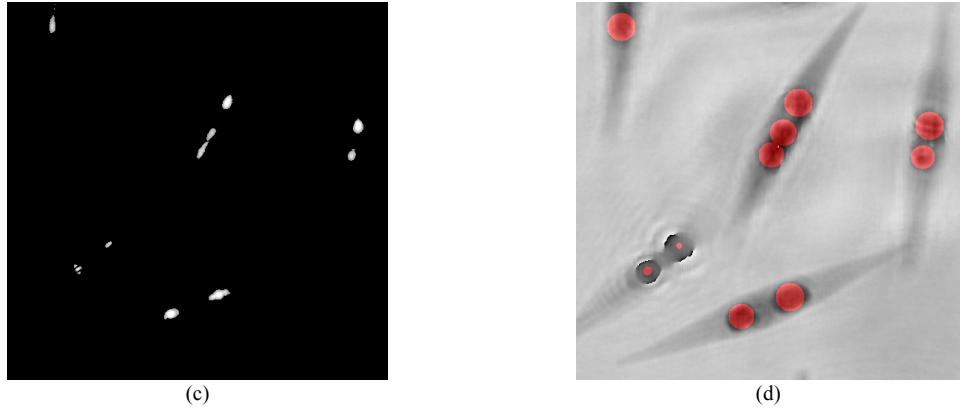

Fig. 3. Validation of the proposed method with correlative quantitative phase contrast–fluorescence imaging. (a) Inverted phase of the recorded hologram [(1), focused cell; (2), (4), and (5), slightly out of focus cells; and (3), largely out of focus cell]; (b) corresponding fluorescence image of the LDs stained with BODIPY 505/515; (c) accumulation of the zones of maximal intensities, which correspond to the intensity peaks created by the LDs; and (d) superimposition of the inverted phase of (a) with the computed disks, in red, which represent the assessed LDs. Scale bar = 5μm.

In Figs. 3(a) and 3(b), it can be seen that two focus fluorescent LDs stained with BODIPY 505/515 inside the cell 1 are clearly identified in the corresponding phase image. Figure 3(c) shows the formation of the intensity peaks by these two LDs, thanks to the numerical post-recording reconstruction in depth of the corresponding hologram. With the developed method, the computed diameters of these two LDs are estimated at 2.1 μm and 2.8 μm, as represented in red in Fig. 3(d), while an estimation performed on the fluorescent image gives, respectively, 2.4 μm and 3.2 μm.

In Fig. 3(a), the three cells numbered (2), (4), and (5) are slightly out of focus, which makes the LD fluorescent emission very weak. Nevertheless, we still obtained the LD focalized intensity peaks for those cells, as seen in Fig. 3(c), and their diameter can be evaluated with our method [Fig. 3(d)]. For the LDs inside the cell (2), we obtained three disks, possibly due to the elongated form of one of the two LDs. It provides however a good approximation of the LD total volume for this cell.

For the largely out of focus cell (3) in Fig. 3(a), the fluorescence emission is too weak to be recorded [Fig. 3(b)]. We obtained the focalized intensity peaks for this cell [Fig. 3(c)] but the LD diameter evaluation is disturbed by their unfocused positions in the hologram, as seen in Fig. 3(d). An individual refocusing of cells with the criterion in Eq. (3), as efficiently demonstrated in [33], would allow a proper evaluation in this case.

We observed that the fluorescent emission of the BODIPY 505/515 suffers from very fast fading. Therefore, for labeled samples, it was experimentally not possible to accurately focus each fluorescent LD, record the fluorescence image, and the corresponding hologram.

A few other images of fluorescent LDs are shown in Fig. 4. Table 2 gives a comparison of the size of the different LDs in Fig. 4 measured by both holographic and fluorescent methods.

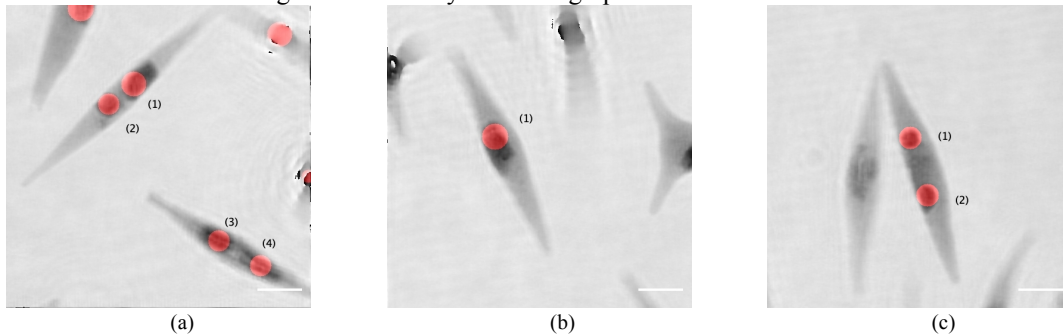

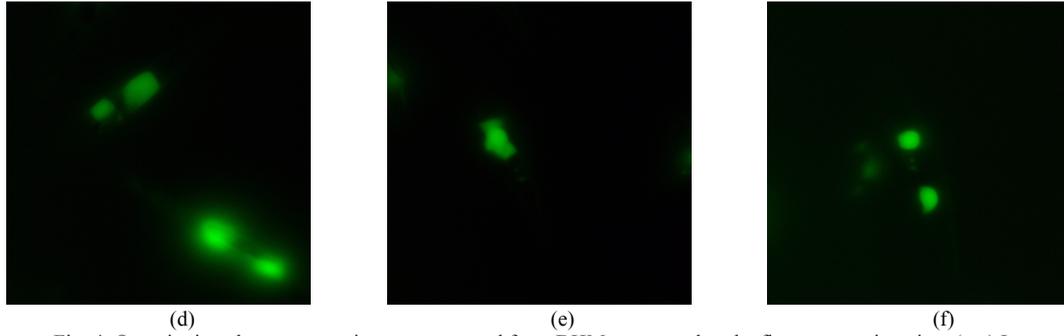

<div align="center">(d)           (e)           (f)</div>

Fig. 4. Quantitative phase contrast images, extracted from DHM, compared to the fluorescence imaging. (a-c) Inverted quantitative phase images superposed with the red disks computed with the proposed method; (d-f) the corresponding fluorescent images with the LDs stained with BODIPY 505/515. Scale bar = 5 µm.

| LD identification in Fig. 4 | Diameters - Holographic method | Estimated diameter - Fluorescence |
|---|---|---|
| Fig. 4(a), LD(1) | 2.9 µm | 3.9 µm |
| Fig. 4(a), LD(2) | 2.8 µm | 2.3 µm |
| Fig. 4(a), LD(3) | 2.4 µm | Not measurable |
| Fig. 4(a), LD(4) | 2.4 µm | Not measurable |
| Fig. 4(b), LD(1) | 2.9 µm | 2.8 µm |
| Fig. 4(c), LD(1) | 2.5 µm | 2.6 µm |
| Fig. 4(c), LD(2) | 2.5 µm | 2.4 µm |

Table 2. Comparison of the Measured LD Size with the Proposed and the Fluorescence Method

In Table 2, except for the Fig. 4(a), LD(1), we can observe that the differences between the mean diameters obtained by the holographic and fluorescent methods are weak. Indeed, we have to compare the obtained values with the resolution limit equal to $0.61\lambda/\text{NA} \approx 0.25\,\mu\text{m}$. We observe that the differences between the fluorescent and holographic methods are smaller than the resolution limit. For the Fig. 4(a), LD(1) case, the deviation is due to the fact that this LD is highly elongated, as observed on both fluorescent and quantitative phase images, that may introduce a significant deviation. However, this kind of situation has a weak impact for the establishment of statistical data to characterize LDs in a culture. We have also to point out that the fluorescent images of Fig. 4(a), LD(3) and Fig. 4(a), LD(4) cannot be measured due to the fact that they are slightly out of focus. This also constitutes a major limitation of the methods based on fluorescent imaging, which is solved by using DHM.

In conclusion, the results obtained for the focus fluorescent LDs confirm a correct evaluation of the LD sizes obtained with the proposed holographic method.

### 3.3. Validation of the method on calibrated particles

In order to validate our method for both detection and size assessment, we applied it to monodisperse silica particles having a mean diameter of 7.82 µm (standard deviation = 0.31 µm). Monitoring such larger particles is useful as this size can be reached for bigger LDs in other cell types such as adipocytes.

To perform the test, we prepared a suspension of these particles in water injected into an ibidi chamber. We then placed the ibidi slide in the DHM equipped with $\times 40$, NA = 0.60, microscope lenses. We recorded 60 images at different locations of the sample by translating the ibidi slide. The sedimented particles are recorded out of focus. The holograms are processed by the Fourier method to extract the amplitude modulus and the phase, as in the case of the LDs. Figure 5(a) shows an example of a recorded amplitude modulus whereas the corresponding phase, in the recording plane, is presented in Fig. 5(b).

The holographic complex amplitude reconstruction is applied by steps of 0.5 µm to obtain the refocused image by using Eqs. (1)–(3), as shown in Figs. 5(c) and (d) (reconstruction at −9.5 µm). As the particles are sedimented, the refocusing criterion can be applied to the full hologram field.

According to our method, the digital holographic reconstruction can be then applied in order to search and detect the maximum peak intensities of each particle [Fig. 5(e)]. The reconstruction step is 0.5 µm. Due to the particle size dispersion, the reconstruction distance to reach the maximum intensity slightly varies for each one. We obtained an average focalization distance *r* of 14 µm.

Knowing the focalization distance *r* with respect to the focus plane, and measuring the diameter *D* in Fig. 5(d) for several particles, it is then possible to evaluate the multiplicative factor *b*. Finally, we compute the size of all the particles of the sequence according to their own maximum focusing peak distance *r*, using the equation $br = D$, with this value of *b*. Figure 5(f) shows the superimposition on the in-focus particles, in grey, of the computed ones, in red, according to the described method. We observe that it is well superimposed.

With the proposed method, we obtained that the average diameter is 8.33 µm with a standard deviation of 0.60 µm. With the NA=0.6, the resolution limit is 0.54µm. The difference between the supplier data and our result is 0.51µm, which is lesser than the resolution limit. Therefore, we can conclude that our results are in good agreement with the data of the particle supplier.

It is important to outline in this example that the proposed method allows to discriminate particles that are in contact, as we can observe in Fig. 5. Indeed, the focused spots by application of the digital holographic propagation [Fig. 5(e)] are well separated, giving rise to the right identification of the particles without having to use further segmentation processes.

It is interesting to note that the method also allows the addition of complementary criteria to select the elements we want to measure. For example, it is possible to reject too small intensity peaks or to eliminate peaks that are too far with respect to the focus plane.

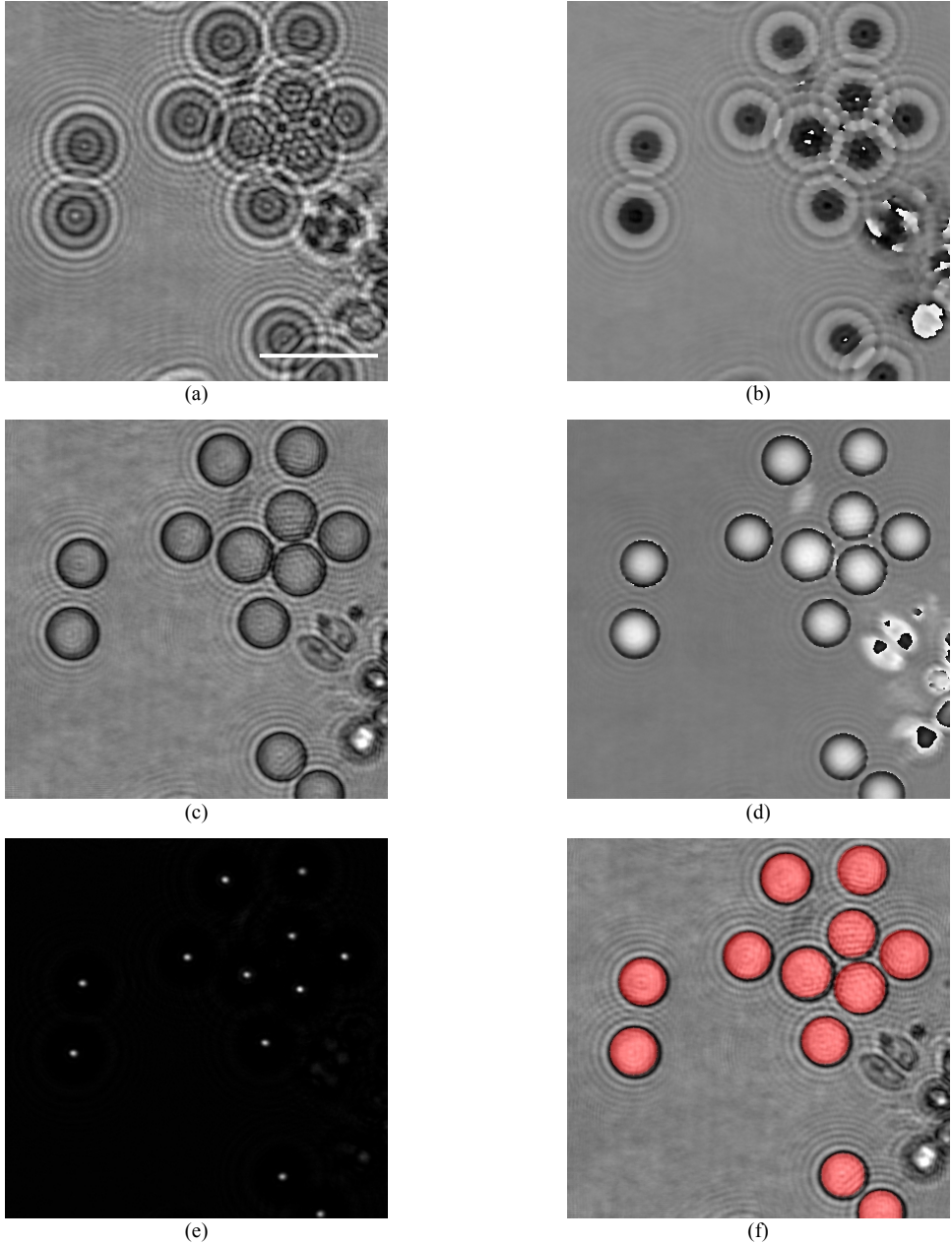

(a)  (b)

(c)  (d)

(e)  (f)

Fig. 5. Validation of the proposed method on larger particles. (a) Amplitude modulus and (b) phase images of monodisperse calibrated silica particles, in the recorded plane; (c) refocused amplitude modulus and (d) phase; (e) reconstruction by digital holography of the intensities over a distance of 15.5 µm to show the intensity peaks created by silica particles; and (f) superimposition on (c) of the disks computed according to the method, in red. Scale bar = 20 µm.

## 4. Discussion and conclusions

Thanks to DHM, we developed a non-destructive method that allows to evaluate the lipid content of *Phaeodactylum tricornutum* living cells in a non-invasive and fast way. The evaluation of the mean value of LD volume is very close to the measurement realized by Wong *et al.* [12] with a laser scanning confocal microscopy on *Phaeodactylum tricornutum* in the same culture conditions.

The developed method was cross-validated thanks to the LD staining with the fluorochrome BODIPY 505/515 and the use of a multimode DHM that allows to perform correlative quantitative phase contrast-fluorescence microscopy. The method was also validated on calibrated and larger silica particles. This is meaningful as larger LDs can be found inside other cell types, as for example adipocytes.

Our method is validated for spherical LDs. This assumption seems acceptable, as we know LDs tend to be spheroid objects. However more accurate shape measurements request more sophisticated and constraining instrumentations and processes as optical diffraction tomography [20], or CARS microscopy [14,15] that make the analysis of a large number of cells more difficult. Fluorescence microscopy [7–10] and Laser scanning confocal microscopy [11,12] are inherently invasive and slow as there is a need to stain and to accurately focus each LD or to scan the sample. Moreover there is the critical fading phenomenon for some fluorochromes as BODIPY 505/515. The proposed technique does not have such constraints and allows a non-invasive and non-destructive analysis of a large number of LDs, leading to a statistical data analysis of evolving cultures more rapidly and efficiently. This aspect is crucial with respect to the actual needs to monitor the microalgal cultures.

It is possible, if necessary, to automatize the *b*-parameter assessment. In this case, a threshold must be used to detect the LDs in a few, randomly chosen, phase images. Among all the areas above this threshold, only those corresponding to one single intensity peak, i.e., one single LD, must be kept. The measurement of the surface area of every LD gives a corresponding diameter *D*, and this diameter *D* is divided by the value of the focusing distance *r* of this LD, providing an assessment of *b* for this LD. Repeating the process for a few randomly-chosen LDs, and considering the measurement errors in *D* and *r*, which are related to the resolution limit, the *b* value is assessed, as well as the relative error $\overline{\Delta b/b}$. Even if such an automatic process would make the parameter estimation easier, which could be valuable, e.g., if many various cultures are studied, it is important to note that the reached accuracy would remain similar. It is also interesting to remind that the method we proposed in this paper provides a tool able to monitor the evolution of LDs during microalgal cultures; therefore, what is important is the evolution of the data computed with the same *b* value, the absolute value of *b* being less relevant in this case.

The method can be applied for the quantification of LDs of other microalgal species such as *Chlamydomonas* sp., *Monoraphidium neglectum*, or *Tetraselmis suecica*, for example, for which the LDs are spheroid. Even in the case of different LDs partially overlapping in the cell, the method should allow to identify the 3D localization of individual LDs inside the cell and to estimate their volume.

The method is quite effective with a short sedimentation step of the culture samples, in such a way that the refocusing of the cells can be performed in full-field images. However, it can be also used for non-sedimented and unfocused microalgal cells. In this case, each LD can be individually refocused, by using for instance the efficient method described in [33].

In order to increase the analysis speed of the lipid microalgal cultures with the developed method, the sample can be laterally translated with a motorized stage and the holograms can be recorded with a faster camera. The microalgal culture samples could also be analyzed for the lipid content in a high throughput way by the DHM that we already developed for the in-flow analysis of plankton microorganisms [33]. In this case, as with the proposed configuration a full hologram is obtained for every snap shot, the hologram can be recorded very fast, with a short exposure time. In a routine configuration, one can expect an exposure time $t_e$ of 10 µs. It is then possible to analyze flux with an acceptable maximum speed equal to the resolution ($0.61\lambda/\mathrm{NA}$) divided by $t_e$, which gives $2.5 \cdot 10^4$ µm/s. Therefore, a large amount of in-flow images per second (typically 100 i/s) can be recorded, to rapidly acquire representative hologram samples of cultures, e.g., within 100 µm thickness flow cells. Moreover, most of the digital processing can be implemented on GPU to considerably reduce the processing time, to achieve fast analysis. Consequently, the reliable method described in this paper could also be implemented in an in-flow DHM, resulting in very rapidly providing the information about the lipid content of the culture.

**Funding.** Fondation Philippe Wiener – Maurice Anspach (FWA, Belgium); Fonds de la Recherche Scientifique – FNRS (F.R.S.-FNRS, Belgium).

**Acknowledgments.** The authors acknowledge Nathalie Gypens, Boris Wittek, and Colin Royer of the Laboratoire d'Ecologie des Systèmes Aquatiques de l'Université libre de Bruxelles for their support for the *Phaeodactylum tricornutum* culture. During this study, Dr Jérôme Dohet-Eraly was first a Postdoctoral Research Fellow of the Fondation Philippe Wiener – Maurice Anspach (FWA, Belgium) and then Chargé de Recherches du Fonds de la Recherche Scientifique – FNRS (F.R.S.-FNRS, Belgium).

**Disclosures.** The authors declare no conflicts of interest.